\documentstyle[aps,preprint]{revtex}
\tightenlines
\begin{document}

\pagestyle{plain}
\setcounter{page}{1}
\setcounter{footnote}{00}

\renewcommand{\thefootnote}{\alph{footnote}}

\baselineskip=18pt 
\def\doublespaced{\baselineskip=\normalbaselineskip\multiply
    \baselineskip by 150\divide\baselineskip by 100}


%
%
\begin{titlepage}
\baselineskip=0.2in
\begin{flushright}
HD-THEP-98-23\\
\end{flushright}
\vspace{0.2in}
\begin{center}
{\large 
   Extracting Top Quark CP Violating Dipole Couplings        
            via  $t\bar t\gamma$ and ~$t\bar tZ$ Productions at the LHC  }\\
\vspace{.2in}
Hong-Yi Zhou\\
\vspace{.2in}
        Institut f\"{u}r Theoretische  Physik, Universit\"{a}t Heidelberg,
        Philosophenweg 16, D-69120 Heidelberg, Germany
{\footnote{Present mailing address}}\\
              and \\
    Institute of Modern Physics and Department of Physics,\\
    Tsinghua University, Beijing 100084, P.R. China\\

\end{center}
\vspace{.3in}

\begin{center}\begin{minipage}{5in}
\baselineskip=0.25in
\begin{center} Abstract\end{center}

We propose to extract the electric and weak dipole moments of 
the top quark via $t\bar t\gamma$ and $t\bar t Z$ productions at the 
CERN LHC. With the large numbers of events available at the 
LHC, these dipole moments can be measured to the accuracy of 
$10^{-18}e~cm$.

\end{minipage}\end{center}

\vspace{.5in}


\end{titlepage}
\newpage

\baselineskip=18pt 
\renewcommand{\thefootnote}{\arabic{footnote}}
\setcounter{footnote}{0}

\section{ Introduction }
\indent
In a recent work\cite{zhy}, we studied on the measurements 
of the top quark electric dipole moment(EDM) and 
weak dipole moment(WDM) via top quark pair production at 
the NLC. These dipole moments are CP violating. It was shown that 
these dipole moments can be measured to the accuracy of 
$10^{-18}e~cm$ at $\sqrt{s}=500$ GeV $e^+e^-$ collider with 
$50fb^{-1}$ integrated luminosity by using  optimal observables. 
The methods of measuring the top quark dipole moments 
via  top quark pair production  at $e^+e^-$ colliders 
have also been studied in \cite{review}-\cite{recent}. 
In this letter, we propose to measure the top quark dipole moments 
at the LHC via $t\bar tV$($V=\gamma,~Z$) production as 
complimentary measurements to those obtained from the NLC.  
At the LHC, one can obtain a detector accumulated integrated 
luminosity of about $600fb^{-1}$ which will result in 
large number of $t\bar t V$ events. So that it may be  possible to 
obtain better limits on the top quark dipole moments at the LHC 
than at the $500$ GeV NLC. 
The processes of $t\bar t V$ productions are somewhat similar to 
$e^+e^-\to Z\to \tau^+\tau^-\gamma$ studied in Ref.\cite{zdecay} for 
measuring $\tau$ dipole moments. In Ref.\cite{zdecay}, it is found that 
naive CP-odd observables are not good for extracting the 
CP violating effects and the optimal observables are more effective. 
Therefore, in this study, we shall apply only the optimal observables 
\cite{opt1}-\cite{opt3} to extract the CP violating effects.

\section{ Calculations}
\indent
The couplings between the top quark and $\gamma$, $Z$ bosons 
take the form:  
\begin{eqnarray}
& &-ie[g_t^V\gamma^\mu(1+\alpha_t^V\gamma_5)+\gamma^\mu\gamma^\nu k^V_{\nu}
(id_t^V/e)\gamma_5],
\end{eqnarray}
where $k^V$ is the outgoing momentum of $V=\gamma$ or $~Z$. 
$d_t^V$ is the dipole moment which we assume to have 
imaginary part as well as real part. 
We denote $\hat{d}_t^V=d_t^V/e$.  The other couplings are:
\begin{eqnarray}
& & g_t^\gamma=2/3,~\alpha_t^\gamma=0,\\
& & g_t^Z=\frac{1-\frac{8}{3}\sin^2\theta_W}{4\sin\theta_W \cos\theta_W}~,
\alpha_t^Z=-\frac{1}{1-\frac{8}{3}\sin^2\theta_W}~.
\end{eqnarray}

At the LHC, the main production process of $t\bar tV$ is 
$gg\to t\bar t V$ which is shown in Fig. 1(the corresponding diagrams 
of Fig.(c)-(e) with the interchanging of the two gluons are not depicted).
We shall assume the dipole moments are small enough that their 
quadratic contributions to the total cross section are negligible. 
Therefore the dipole moments contribute only to the CP violating 
effects through their interference with the standard model(SM)  
contribution. This interference   is linear in $\hat{d}_t^V$. 
To observe the CP violating effects, one needs 
to know the spins of the top quarks which can be determined 
statistically from their decay products. 
We assume the SM decay of the top quark and apply the narrow width 
approximations of the top quark and W-boson propagators:
\begin{eqnarray}                  
\frac{1}{|q^2_X-m_X^2+im_X\Gamma_X|^2}\rightarrow 
\frac{\pi}{m_X\Gamma_X}\delta(q_X^2-m_X^2)~,
\end{eqnarray} 
where $X$ stands for top quark and W-boson, $\Gamma_X$ is the width of 
$X$.  

The  cross section for reaction 
$pp \to t\bar tV\to bl_1^+\nu_{l_1}\bar b l_2^-{\bar\nu}_{l_2}V$ 
($b\bar q_1 q_1' \bar b q_2\bar q'_2V$) can be written as
\begin{eqnarray} 
\label{cross}
d\sigma&=& \frac{f_p^g(x_1)f_p^g(x_2)}{2\hat{s}(8\pi)^{8}}
\frac{\lambda_t |M_D|^2dx_1dx_2 }
{m_t^2m_W^2\Gamma_t^2\Gamma_W^2}d\Phi_{t\bar tV} d\Omega_{W^+}' d\Omega_{W^-}'
d\Omega_{l_1^+}'d\Omega_{l_2^-}' ~,   
\end{eqnarray} 
where $f_p^g(x)$ is the gluon distribution function in proton. 
$\hat{s}$ and $d\Phi_{t\bar tV}$ are the C.M. energy 
and phase space element of the subprocess $gg\to t\bar tV$,respectively, 
and 
\begin{eqnarray} 
\lambda_t=(1-\frac{(m_W+m_b)^2}
{m_t^2})(1-\frac{(m_W-m_b)^2}{m_t^2})\approx (m_t^2-m_W^2)^2/m_t^4~,
\end{eqnarray} 
$d\Omega_{W^+}' (d\Omega_{W^-}')$ is the solid angle element of $W^+(W^-)$
in the rest frame of the (anti) top quark,  
$d\Omega_{l_1^+}'(d\Omega_{l_2^-}')$ denotes the     
solid angle element of $l_1^+(l_2^-)$ in the rest frame of $W^+(W^-)$, 
$|M_D|^2$ is the amplitude  square excluding the top quark and W-boson 
propagators after the decays of the top quarks.  

In our calculations, $|M_D|^2$ is easily 
obtained from the amplitude of $gg\to t\bar tV$ by the following 
substitutions:
\begin{eqnarray} 
& & \bar u(p_t)\rightarrow \frac{g^2}{8}\bar u_b \gamma_\mu(1-\gamma_5)
(\rlap/p_t+m_t)\bar u_{\nu_1}\gamma^\mu(1-\gamma_5) v_{l_1}~,\\\nonumber
& & v(p_{\bar t})\rightarrow \frac{g^2}{8}\bar u_{l_2} \gamma_\mu(1-\gamma_5)
v_{\nu_2}(\rlap/p_{\bar t}-m_t)\gamma^\mu(1-\gamma_5) v_{\bar b}~,
\end{eqnarray} 
where $g$ is the weak $SU(2)$ coupling constant and 
$\bar u,~v$ are Fermion wave functions. 
The above expresssions  
are calculated numerically. Denoting $g_1^V=Re(\hat{d}_t^V)$ and 
$g_2^V=Im(\hat{d}_t^V)$, we can write $|M_D|^2$ as  
\begin{eqnarray}  
|M_D|^2 &=&\Sigma_0+g_1^V\Sigma_{1}+g_2^V\Sigma_{2},
\end{eqnarray} 
where $\Sigma_0$ is the SM amplitude square.$\Sigma_{1}$ and 
$\Sigma_{2}$ are CP-odd amplitude terms which do not contribute 
to total cross section.  

The optimized CP-odd observables in the full final state 
phase space  are defined by 
\begin{eqnarray}
O_{11}=\frac{\Sigma_{1}}{\Sigma_0}~,
O_{12}=\frac{\Sigma_{2}}{\Sigma_0}~. 
\end{eqnarray} 

When the top quark decays hadronically, we can not distinguish 
quark and antiquark jet. For hadronic-leptonic 
events, the missing neutrino momenta can be fully reconstructed 
using energy momentum conservation equations, so that we are 
left with two fold ambiguity of the jet momenta. For purely 
hadronic events, we have four fold ambiguity.     
Considering this ambiguity, one can 
define alternatively the optimal observables:
\begin{eqnarray}
O_{2i}=\frac{\sum\limits_{j}\Sigma_i}{\sum\limits_{j}\Sigma_0},~~~ 
O_{4i}=\frac{\sum\limits_{j'}\Sigma_i}{\sum\limits_{j'}\Sigma_0}, 
\end{eqnarray} 
where $i=1,2$ and the sum  $j$ is over the two possible assignments of the 
jet momenta to the quark and antiquark in hadronic-leptonic events.
$j'$ is over the possible assignments of the 
jet momenta to the quark and antiquark in purely hadronic events.  
Because the number of events for two top quarks decay  
semi-leptonically is small, we shall not consider this case.  

The mean value of the observable $O_{2i}$ is defined as 
\begin{eqnarray} 
\langle O_{2i}\rangle&=&
\frac{\int d\sigma^+ O_{2i}^++   
\int d\sigma^- O_{2i}^-}{\int d\sigma^++\int d\sigma^-}~,   
\end{eqnarray} 
where the superscript $+,-$ mean that the integrations are over 
$bl_1^+\nu_{l_1}\bar b q_2\bar q'_2V$ and 
$b\bar q_1 q_1'\bar b l_2^-{\bar\nu}_{l_2}V$ final states, respectively.  
The mean value of the observable $O_{4i}$ is simply  
\begin{eqnarray} 
\langle O_{4i}\rangle&=&
\frac{\int d\sigma O_{4i}}{\int d\sigma}~.   
\end{eqnarray} 

We can express the mean values by 
\begin{eqnarray}
& &\langle O_{ni}\rangle= c_{ni}g_i^V,~~  
\end{eqnarray} 
where $n=2,4$ and $c_{ni}=\langle O_{ni}\rangle ^2$. 
The $1\sigma$ level($68\%$ C.L.) 
statistical error of $g_i^V$  is 
given by
\begin{eqnarray}
\label{err}
\Delta g_{ni}^V=\frac{1}{\sqrt{Nc_{ni}}},~~ 
\end{eqnarray}
where $N$ is the number of events. To reduce the statistcal errors, 
one can combine the measurements of $O_{2i}$ and $O_{4i}$ to get 
a combined error $\Delta g_{ci}^V$\cite{data}:
\begin{eqnarray}
\frac{1}{(\Delta g_{ci}^V)^2}=\frac{1}{(\Delta g_{2i}^V)^2}+
\frac{1}{(\Delta g_{4i}^V)^2}, 
\end{eqnarray}
where $\Delta g_{2i}^V$ and $\Delta g_{4i}^V$ are the errors of the 
measurements using $O_{2i}$ and $O_{4i}$,respectively.

\section{Results}

To calculate the number of events, 
we  have  used the following parameters in our 
calculations: (1) the overall detection efficiency 
 $\epsilon=0.1$; (2) the integrated luminosity  ${\cal L}=600fb^{-1}$;
(3) the branching ratio of hadronic-leptonic top quark decays 
$B_{lj}^t=0.29$($l=e,\mu$),the branching ratio of 
the purely hadronic top quark decays $B_{jj}^t=0.46$; 
(4) the branching ratio of leptonic Z decay $B^Z_{ll}=0.067$.  
We consider only the leptonic Z decay channel in the following 
calculations. The number of events is given by 
\begin{eqnarray}
N&=&\epsilon{\cal L}\sigma B^t B^V,
\end{eqnarray}
where $\sigma$ is the total $t\bar tV$ production cross section, 
$B^t=B_{lj}^t$ or $ B_{jj}^t$,$B^\gamma=1,~B^Z=B^Z_{ll}$. 

In calculating the total cross sections and 
the mean values of the observables, we set $m_t=176$ GeV,
$\alpha_{em}=1/128.8$  and 
$pp$ energy $\sqrt{s}=14$TeV.   
The parton distribution functions  of MRS A' are used with 
$Q^2=m_t^2$\cite{MRSA}. In order to give more realistic results,
we also apply the following cuts on $t, \bar t, V$:
\begin{eqnarray}  
p_T(i)>20 GeV~,~~ |y(i)|<2.5,~,~~\Delta R(i,j)>0.4~,
\end{eqnarray}
where $i,~j=t,~\bar t,~V$, $p_T$ is the transverse momentum, 
$y$ is the rapidity and $\Delta R=\sqrt{(\Delta y)^2+(\Delta\phi)^2}$ 
is the solid angle separation of two particles. 

With the above parameters and cut conditions, we get the following 
results for the total cross sections 
\begin{eqnarray}
 \sigma(t\bar t\gamma)=1.465 pb,~~
 \sigma(t\bar tZ)=0.614 pb. 
\end{eqnarray}

In Table I., we present the $1\sigma$ statistical errors of 
the dipole couplings. The results for $\hat{d}_t^\gamma$ are 
even better than the $500$ GeV NLC with $50fb^{-1}$\cite{zhy}.
Due to the small branching ratio of $Z$ decay to leptons, the 
results for $\hat{d}_t^Z$ are not as good as that in Ref.\cite{zhy}. 
Our conclusion is that the limits on the top 
quark dipole couplings which can obtained from $t\bar tV$ 
production at the LHC with $600fb^{-1}$ are about $10^{-18} e~cm$.
These limits are comparable with those obtained from 
$e^+e^-\to t\bar t$ at the 500 GeV NLC.

\begin{table}
\caption{} 
$1\sigma$ statistical errors of the coupling constants $g_i^V$.
Unit:$10^{-18}~cm$.   
\small{
\begin{center}
\begin{tabular}{|c|cccc|}
& $g_{1}^\gamma$ & $g_{2}^\gamma$ & $g_{1}^Z$ & $g_{2}^Z$   \\ \hline
$O_{2i}$ &3.4 & 1.1 & 10.4 & 6.0 \\
$O_{4i}$ &3.5 & 1.2 & 10.6 & 5.0 \\
combined &2.4 & 0.8 & 7.4 & 3.9 \\ 
\end{tabular}
\end{center}
}

\end{table}

\begin{center}
{\bf Acknowledgements} 
\end{center}

The author is financially supported by the Alexander von Humboldt  
Foundation of Germany.

\vspace{5cm}

\newpage 
\begin{figure}[h]

\vspace*{20cm}
\includegraphics{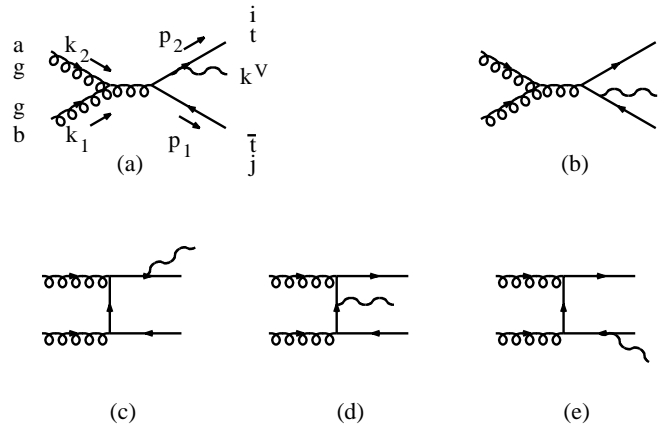}

\vspace{-1.0cm}

\caption[]{
Feynman diagrams of $gg\to t\bar t V$($V=\gamma,Z$).}
\label{frames}
\end{figure}

\end{document}